\documentclass[runningheads,a4paper]{llncs}
\usepackage{amsmath}
\usepackage{graphicx}
\usepackage{amssymb}
\usepackage{comment} 
\usepackage{url}
\let\citep=\cite

\newcommand{\keywords}[1]{\par\addvspace\baselineskip
\noindent\keywordname\enspace\ignorespaces#1}
\newcommand{\res}[1]{{\footnotesize \sffamily \scshape #1}}
\begin{document}
\mainmatter  

\title{An Inductive Logic Programming Approach to Validate Hexose
  Binding Biochemical Knowledge}

\author{Houssam Nassif\inst{1,2} \and Hassan Al-Ali\inst{3} \and
  Sawsan Khuri\inst{4,5} \and Walid Keirouz\inst{6} \and David Page\inst{1,2}}

\authorrunning{An ILP Approach to Validate Hexose Binding Biochemical
  Knowledge}

\institute{Department of Computer Sciences,
\and
Department of Biostatistics and Medical Informatics, 
\\University of Wisconsin-Madison, USA
\and
Department of Biochemistry and Molecular Biology, 
\and
Center for Computational Science, University of Miami,
\and
The Dr. John T. Macdonald Foundation Department of Human Genetics, 
\\University of Miami Miller School of Medicine, Florida, USA
\and
Department of Computer Science, American University of Beirut, Lebanon}

\maketitle

\begin{abstract}
  Hexoses are simple sugars that play a key role in many cellular
  pathways, and in the regulation of development and disease
  mechanisms.  Current protein-sugar computational models are based,
  at least partially, on prior biochemical findings and knowledge.
  They incorporate different parts of these findings in predictive
  black-box models.  We investigate the empirical support for
  biochemical findings by comparing Inductive Logic Programming (ILP)
  induced rules to actual biochemical results.  We mine the Protein
  Data Bank for a representative data set of hexose binding sites,
  non-hexose binding sites and surface grooves.  We build an ILP model
  of hexose-binding sites and evaluate our results against several
  baseline machine learning classifiers.  Our method achieves an
  accuracy similar to that of other black-box classifiers while
  providing insight into the discriminating process. In addition, it
  confirms wet-lab findings and reveals a previously unreported
  \res{Trp-Glu} amino acids dependency.

\end{abstract}

\section{Introduction}
Inductive Logic Programming (ILP) has been shown to perform well in
predicting various substrate-protein bindings
(e.g.,~\cite{FINN98,srinivasan97}).  In this paper we apply ILP to a
different and well studied binding task.

Hexoses are $6$-carbon simple sugar molecules that play a key role in
different biochemical pathways, including cellular energy release,
signaling, carbohydrate synthesis, and the regulation of gene
expression~\citep{BIOL201}.  Hexose binding proteins belong to diverse
functional families that lack significant sequence or, often,
structural similarity~\citep{Khuri1}.  Despite this fact, these
proteins show high specificity to their hexose ligands.  The few amino
acids (also called residues) present at the binding site play a large
role in determining the binding site's distinctive topology and
biochemical properties and hence the ligand type and the protein's
functionality.

Wet-lab experiments discover hexose-protein properties.  Computational
hexose classifiers incorporate different parts of these findings in
black-box models as the base of prediction.  No work to date has taken
the opposite approach: given hexose binding sites data, what
biochemical rules can we extract with no prior biochemical knowledge,
and what is the performance of the resulting classifier based solely
on the extracted rules?

This work presents an ILP classifier that extracts rules from the data
without prior biochemical knowledge.  It classifies binding sites
based on the extracted biochemical rules, clearly specifying the rules
used to discriminate each instance.  Rule learning is especially
appealing because of its easy-to-understand format.  A set of if-then
rules describing a certain concept is highly expressive and
readable~\citep{Mitchell}.  We evaluate our results against several
baseline machine learning classifiers.  This inductive data-driven
approach validates the biochemical findings and allows a better
understanding of the black-box classifiers' output.

\section{Previous Work}
Although no previous work tackled data-driven rule generation or validation, many
researchers studied hexose binding.

\subsection{Biochemical Findings}
From the biochemical perspective, Rao et al.~\cite{Rao1} fully
characterized the architecture of sugar binding in the Lectin protein
family and identified conserved loop structures as essential for sugar
recognition.  Later, Quiocho and Vyas~\cite{Quiocho} presented a
review of the biochemical characteristics of carbohydrate binding
sites and identified the planar polar residues (\res{Asn, Asp, Gln,
  Glu, Arg}) as the most frequently involved residues in hydrogen
bonding.  They also found that the aromatic residues \res{Trp, Tyr},
and \res{Phe}, as well as \res{His}, stack against the apolar surface
of the sugar pyranose ring.  Quiocho and Vyas also pinpointed the role
of metal ions in determining substrate specificity and affinity.
Ordered water molecules bound to protein surfaces are also involved in
protein-ligand interaction~\citep{Water}.

Taroni et al.~\cite{Taroni} analyzed the characteristic properties of
sugar binding sites and described a residue propensity parameter that
best discriminates sugar binding sites from other protein-surface
patches.  They also note that simple sugars typically have a
hydrophilic side group which establishes hydrogen bonds and a
hydrophobic core that is able to stack against aromatic residues.
Sugar binding sites are thus neither strictly hydrophobic nor strictly
hydrophilic, due to the dual nature of sugar docking.  In fact, as
Garc\'{i}a-Hern\'{a}ndez et al.~\cite{Garcia} showed, some polar groups in
the protein-carbohydrate complex behave hydrophobically.

\subsection{Computational Models}
Some of this biochemical information has been used in computational
work with the objective of accurately predicting sugar binding sites
in proteins.  Taroni et al.~\cite{Taroni} devised a probability
formula by combining individual attribute scores.  Shionyu-Mitsuyama
et al.~\cite{Shionyu} used atom type densities within binding sites to
develop an algorithm for predicting carbohydrate binding.  Chakrabarti
et al.~\cite{Chakrabarti} modeled one glucose binding site and one
galactose binding site by optimizing their binding affinity under
geometric and folding free energy constraints.  Other researchers
formulated a signature for characterizing galactose binding sites
based on geometric constraints, pyranose ring proximity and hydrogen
bonding atoms~\citep{SUJ1,SUJ2}.  They implemented a 3D structure
searching algorithm, COTRAN, to identify galactose binding sites.

More recently, researchers used machine learning algorithms to model
hexose binding sites.  Malik and Ahmad~\cite{Malik} used a Neural
Network to predict general carbohydrate as well as specific galactose
binding sites.  Nassif et al.~\cite{Nassif} used Support Vector
Machines to model and predict glucose binding sites in a wide range of
proteins.

\section{Data Set}
The Protein Data Bank (PDB)~\citep{PDB1} is the largest repository of
experimentally determined and hypothetical three-dimensional
structures of biological macromolecules.  We mine it for proteins
crystallized with the most common hexoses: galactose, glucose and
mannose~\citep{OrganicChem}.  We ignore theoretical structures and
files older than PDB format $2.1$.  We eliminate redundant structures
using PISCES~\citep{PISCES} with a $30\%$ overall sequence identity
cut-off. We use Swiss-PDBViewer~\citep{SWISSPDB} to detect and discard
sites that are glycosylated or within close proximity to other
ligands. We check the literature to ensure that no hexose-binding site
also binds non-hexoses. The final outcome is a non-redundant positive
data set of $80$ protein-hexose binding sites
(Table~\ref{t:positives}).

\begin{table}[h!]
\caption{Inventory of the hexose-binding positive data set}
\label{t:positives}
\begin{tabular*}{\textwidth}{@{\extracolsep{\fill}}lll*{2}{@{\hspace{22pt}}ll}}
  \hline\noalign{\smallskip}
  Hexose & PDB ID & Ligand & PDB ID & Ligand & PDB ID & Ligand\\
  \noalign{\smallskip}\hline\noalign{\smallskip}
  Glucose & 1BDG & GLC-501 &  1ISY & GLC-1471 &   1SZ2 & BGC-1001\\
          & 1EX1 & GLC-617 &  1J0Y & GLC-1601 &   1SZ2 & BGC-2001\\
          & 1GJW & GLC-701 &  1JG9 & GLC-2000 &   1U2S & GLC-1\\
          & 1GWW & GLC-1371 & 1K1W & GLC-653 &   1UA4 & GLC-1457\\
          & 1H5U & GLC-998 &  1KME & GLC-501 &   1V2B & AGC-1203\\
          & 1HIZ & GLC-1381 &  1MMU & GLC-1 &   1WOQ & GLC-290\\
          & 1HIZ & GLC-1382 &  1NF5 & GLC-125 &   1Z8D & GLC-901\\
          & 1HKC & GLC-915 &  1NSZ & GLC-1400 &   2BQP & GLC-337\\
          & 1HSJ & GLC-671 &  1PWB & GLC-405 &   2BVW & GLC-602\\
          & 1HSJ & GLC-672 &  1Q33 & GLC-400 &   2BVW & GLC-603\\
          & 1I8A & GLC-189 &  1RYD & GLC-601 &   2F2E & AGC-401\\
          & 1ISY & GLC-1461 &  1S5M & AGC-1001 & \\
  \noalign{\smallskip}
  Galactose&1AXZ & GLA-401 &  1MUQ & GAL-301  & 1R47 & GAL-1101\\
           &1DIW & GAL-1400 & 1NS0 & GAL-1400 & 1S5D & GAL-704\\
           &1DJR & GAL-1104 & 1NS2 & GAL-1400 & 1S5E & GAL-751\\
           &1DZQ & GAL-502 &  1NS8 & GAL-1400 & 1S5F & GAL-104\\
           &1EUU & GAL-2 &    1NSM & GAL-1400 & 1SO0 & GAL-500\\
           &1ISZ & GAL-461 &  1NSU & GAL-1400 & 1TLG & GAL-1\\
           &1ISZ & GAL-471 &  1NSX & GAL-1400 & 1UAS & GAL-1501\\
           &1JZ7 & GAL-2001 & 1OKO & GLB-901 &  1UGW & GAL-200\\
           &1KWK & GAL-701 &  1OQL & GAL-265 &  1XC6 & GAL-9011\\
           &1L7K & GAL-500 &  1OQL & GAL-267 &  1ZHJ & GAL-1\\
           &1LTI & GAL-104  & 1PIE & GAL-1  &   2GAL & GAL-998\\
  \noalign{\smallskip}
  Mannose & 1BQP & MAN-402  & 1KZB & MAN-1501 & 1OUR & MAN-301\\
          & 1KLF & MAN-1500 & 1KZC & MAN-1001 & 1QMO & MAN-302\\
          & 1KX1 & MAN-20   & 1KZE & MAN-1001 & 1U4J & MAN-1008\\
          & 1KZA & MAN-1001 & 1OP3 & MAN-503 & 1U4J & MAN-1009\\
  \noalign{\smallskip}\hline
\end{tabular*}
\end{table}


We also extract an equal number of negative examples.  The negative
set is composed of non-hexose binding sites and of non-binding surface
grooves.  We choose $22$ binding-sites that bind hexose-like ligands:
hexose or fructose derivatives, $6$-carbon molecules, and molecules
similar in shape to hexoses (Table~\ref{t:negative_sites}).  We also
select $27$ other-ligand binding sites, ligands who are bigger or
smaller than hexoses (Table~\ref{t:negative_sites}).  Finally, we
specify $31$ non-binding sites: protein surface grooves that look like
binding-sites but are not known to bind any ligand
(Table~\ref{t:negative_random}).

We use $10$-folds cross-validation to train, test and validate our
approach.  We divide the data set in $10$ stratified folds, thus
preserving the proportions of the original set labels and sub-groups.

\begin{table}
\caption{Inventory of the non-hexose-binding negative data set}
\label{t:negative_sites}
\begin{tabular*}{\textwidth}{@{\extracolsep{\fill}}lll@{\hspace{22pt}}lll}
  \hline\noalign{\smallskip}
  PDB ID & Cavity Center & Ligand & PDB ID & Cavity Center & Ligand \\
  \noalign{\smallskip}\hline\noalign{\smallskip}
  \multicolumn{6}{l}{Hexose-like ligands}\\
  \noalign{\smallskip}
  1A8U &  4320, 4323 & BEZ-1 & 1AI7 &  6074, 6077 & IPH-1\\
  1AWB &  4175, 4178 & IPD-2 & 1DBN & pyranose ring & GAL-102\\
  1EOB &  3532, 3536 & DHB-999 & 1F9G &  5792, 5785, 5786 & ASC-950\\
  1G0H &  4045, 4048 & IPD-292 & 1JU4  & 4356, 4359 & BEZ-1\\
  1LBX &  3941, 3944 & IPD-295 & 1LBY &  3944, 3939, 3941 & F6P-295\\
  1LIU &  15441, 15436, 15438 & FBP-580 & 1MOR & pyranose ring & G6P-609\\
  1NCW &  3406, 3409 & BEZ-601 & 1P5D & pyranose ring & G1P-658\\
  1T10 &  4366, 4361, 4363 & F6P-1001 & 1U0F & pyranose ring & G6P-900\\
  1UKB &  2144, 2147 & BEZ-1300 & 1X9I & pyranose ring & G6Q-600\\
  1Y9G &  4124, 4116, 4117 & FRU-801 & 2B0C & pyranose ring & G1P-496\\
  2B32 &  3941, 3944 & IPH-401 &4PBG & pyranose ring & BGP-469\\
  \noalign{\medskip}
  \multicolumn{6}{l}{Other ligands}\\
  \noalign{\smallskip}
  11AS & 5132 & ASN-1  & 11GS & 1672, 1675 & MES-3\\
  1A0J & 6985 & BEN-246 & 1A42 & 2054, 2055 & BZO-555\\
  1A50 & 4939, 4940 & FIP-270 & 1A53 & 2016, 2017 & IGP-300 \\
  1AA1 & 4472, 4474 & 3PG-477 & 1AJN & 6074, 6079 & AAN-1\\
  1AJS & 3276, 3281 & PLA-415 & 1AL8 & 2652 & FMN-360\\
  1B8A & 7224 & ATP-500 & 1BO5 & 7811 & GOL-601\\
  1BOB & 2566 & ACO-400 & 1D09 & 7246 & PAL-1311\\
  1EQY & 3831 & ATP-380 & 1IOL & 2674, 2675 & EST-400\\
  1JTV & 2136, 2137 & TES-500 & 1KF6 & 16674, 16675 & OAA-702\\
  1RTK & 3787, 3784 & GBS-300 & 1TJ4 & 1947 & SUC-1\\
  1TVO & 2857 & FRZ-1001 & 1UK6 & 2142 & PPI-1300\\
  1W8N & 4573, 4585 & DAN-1649 & 1ZYU & 1284, 1286 & SKM-401\\
  2D7S & 3787 & GLU-1008 & 2GAM & 11955 & NGA-502\\
  3PCB & 3421, 3424 & 3HB-550 & & & \\
  \noalign{\smallskip}\hline
\end{tabular*}
\end{table}

\begin{table}
\caption{Inventory of the non-binding surface groove negative data
  set}
\label{t:negative_random}
\begin{tabular*}{\textwidth}{@{\extracolsep{\fill}}ll*{2}{@{\hspace{24pt}}ll}}
  \hline\noalign{\smallskip}
  PDB ID & Cavity Center & PDB ID & Cavity Center & PDB ID & Cavity Center \\
  \noalign{\smallskip}\hline\noalign{\smallskip}
  1A04 &  1424, 2671 & 1A0I &  1689, 799 & 1A22 &  2927\\
  1AA7  & 579 & 1AF7  & 631, 1492 & 1AM2 &  1277\\
  1ARO &  154, 1663 & 1ATG &  1751 & 1C3G  & 630, 888\\
  1C3P &  1089, 1576  & 1DXJ &  867, 1498 & 1EVT &  2149, 2229\\
  1FI2 &  1493 & 1KLM & 4373, 4113 & 1KWP &  1212\\
  1QZ7 &  3592, 2509 & 1YQZ &  4458, 4269 & 1YVB &  1546, 1814\\
  1ZT9 &  1056, 1188 & 2A1K &  2758, 3345 & 2AUP & 2246\\
  2BG9 & 14076, 8076 & 2C9Q & 777 & 2CL3 & 123, 948\\
  2DN2 & 749, 1006 & 2F1K & 316, 642 & 2G50 & 26265, 31672\\
  2G69 & 248, 378 & 2GRK & 369, 380 & 2GSE & 337, 10618\\
  2GSH & 6260 & & & & \\
  \noalign{\smallskip}\hline
\end{tabular*}
\end{table}

\section{Problem Representation}
In this work, we first extract multiple chemical and spatial features
from the binding site. We then apply ILP to generate rules and
classify our data set.

\subsection{Binding Site Representation}
We view the binding site as a sphere centered at the ligand.  We
compute the center of the hexose-binding site as the centroid of the
coordinates of the hexose pyranose ring's six atoms.  For negative
sites, we use the center of the cavity or the ligand's central point.
The farthest pyranose-ring atom from the ring's centroid is located
\mbox{$2.9$ \AA}\ away.  Bobadilla et al.~\cite{BOB1} consider atomic
interactions to be significant within a \mbox{$7$ \AA}\ range.  We
thereby fix the binding site sphere radius to \mbox{$10$ \AA}.  Given
the molecule and the binding site centroid, we extract all atoms
within the sphere.  We include water molecules and ions present in the
binding groove~\citep{Water,Nassif,Quiocho}.  We discard hydrogen
atoms since most PDB entries lack them. We do not extract residues.

For every extracted atom we record its PDB-coordinates, its charge,
hydrogen bonding, and hydrophobicity properties, and its atomic
element and name.  Every PDB file has orthogonal coordinates and all
atom positions are recorded accordingly.  We compute atomic properties
as done by Nassif et al.~\cite{Nassif}.  The partial charge measure
per atom is positive, neutral, or negative; atoms can form hydrogen
bonds or not; hydrophobicity measures are considered as hydrophobic,
hydroneutral, or hydrophilic.  Finally, every PDB-atom has an atomic
element and a specific name. For example, the residue histidine
(\res{His}) has a particular Nitrogen atom named \res{ND1}.  This
atom's element is Nitrogen, and name is \res{ND1}.  Since \res{ND1}
atoms only occur in \res{His} residues, recording atomic names leaks
information about their residues.

\subsection{Aleph Settings}
We use the ILP engine Aleph~\citep{Aleph} to learn first-order rules.
We run Aleph within Yap Prolog~\citep{Yap}.  To speed the search, we
use Aleph's heuristic search.  We estimate the classifier's
performance using $10$-fold cross-validation.

We limit Aleph's running time by restricting the clause length to a
maximum of $8$ literals, with only one in the head.  We set the Aleph
parameter $explore$ to true, so that it will return all
optimal-scoring clauses, rather than a single one, in a case of a tie.
The consequent of any rule is $bind(+site)$, where $site$ is predicted
to be a hexose binding site. No literal can contain terms pertaining
to different binding sites.  As a result, $site$ is the same in all
literals in a clause.

The literal describing the binding site center is:
\begin{equation}
point(+site,-id,-X,-Y,-Z)
\end{equation}
where $site$ is the binding site and $id$ is the binding center's
unique identifier.  $X$, $Y$, and $Z$ specify the PDB-Cartesian
coordinates of the binding site's centroid.

Literals describing individual PDB-atoms are of the form:
\begin{equation}
point(+site,-id,-X,-Y,-Z,-charge,-hbond,-hydro,-elem,-name)
\end{equation}
where $site$ is the binding site and $id$ is the individual atom's
unique identifier.  $X$, $Y$, and $Z$ specify the PDB-Cartesian
coordinates of the atom.  $charge$ is the partial charge, $hbond$
the hydrogen-bonding, and $hydro$ the hydrophobicity.  Lastly,
$elem$ and $name$ refer to the atomic element and its name (see
last paragraph of previous section).

Clause bodies can also use distance literals:
\begin{equation}
dist(+site,+id,+id,\#distance,\#error)\,.
\end{equation}
The $dist$ predicate, depending on usage, either computes or checks
the $distance$ between two points. $site$ is the binding site and the
$id$s are two unique point identifiers (two PDB-atoms or one PDB-atom
and one center).  $distance$ is their Euclidean distance apart and
$error$ the tolerated distance error, resulting in a matching interval
of $distance \pm error$.  We set $error$ to \mbox{$0.5$ \AA}.

We want our rules to refer to properties of PDB-atoms, such as \lq\lq
an atom's name is \res{ND1}\rq\rq, or \lq\lq an atom's charge is not
positive\rq\rq. Syntactically we do this by relating PDB-atoms'
variables to constants using \lq\lq equal\rq\rq\ and \lq\lq not
equal\rq\rq\ literals:
\begin{align}
  &equal(+setting,\#setting)\,,\\
  &not\_equal(+feature,\#feature)\,.
\end{align}
$feature$ is the atomic features $charge$, $hbond$ and $hydro$. In
addition to these atomic features, $setting$ includes $elem$ and
$name$.

Aleph keeps learning rules until it has covered all the training
positive set, and then it labels a test example as positive if
\emph{any} of the rules cover that example.  This has been noted in
previous publications to produce a tendency toward giving more false
positives~\citep{Davis05A,Davis05B}.  To limit our false positives
count, we restrict coverage to a maximum of $5$ training-set
negatives.  Since our approach seeks to validate biological knowledge,
we aim for high precision rules.  Restricting negative rule coverage
also biases generated rules towards high precision.

\section{Results}\label{S:AlephResuts}
The Aleph testing set error averaged to $32.5\%$ with a standard
deviation of $10.54\%$.  The confidence interval is
$[24.97\%,40.03\%]$ at the $95\%$ confidence level.  Refer to
Table~\ref{t:baseline} for the $10$-folds cross-validation accuracies.

To generate the final set of rules, we run Aleph over the whole data
set.  We discard rules with $pos\_cover - neg\_cover \leq 2$.  Even
though Aleph was only looking at atoms, valuable information regarding
amino acids can be inferred.  For example \res{ND1} atoms are only
present within the amino acid \res{His}, and a rule requiring the
presence of \res{ND1} is actually requiring \res{His}.  We present the
rules' biochemical translation while replacing specific atoms by the
amino acids they imply. The queried site is considered hexose binding
if any of these rules apply:

\begin{enumerate}
\item It contains a \res{Trp} residue and a \res{Glu} with an
  \res{OE1} Oxygen atom that is \mbox{$8.53$ \AA}\ away from an Oxygen atom
  with a negative partial charge (\res{Glu, Asp} amino acids, Sulfate,
  Phosphate, residue C-terminus Oxygen).\\ 
  $[$Pos cover = 22, Neg cover = 4$]$\\
\item It contains a \res{Trp}, \res{Phe} or \res{Tyr} residue, an
  \res{Asp} and an \res{Asn}. \res{Asp} and an \res{Asn}'s \res{OD1}
  Oxygen atoms are \mbox{$5.24$ \AA}\ apart.\\ 
  $[$Pos cover = 21, Neg cover = 3$]$\\
\item It contains a \res{Val} or \res{Ile} residue, an \res{Asp} and
  an \res{Asn}.  \res{Asp} and \res{Asn}'s \res{OD1} Oxygen atoms are 
  \mbox{$3.41$ \AA}\ apart.\\ 
  $[$Pos cover = 15, Neg cover = 0$]$\\
\item It contains a hydrophilic non-hydrogen bonding Nitrogen atom
  (\res{Pro, Arg}) with a distance of \mbox{$7.95$ \AA}\ away from a
  \res{His}'s \res{ND1} Nitrogen atom, and \mbox{$9.60$ \AA}\ away from a
  \res{Val} or \res{Ile}'s \res{CG1} Carbon atom.\\ 
  $[$Pos cover = 10, Neg cover = 0$]$\\
\item It has a hydrophobic \res{CD2} Carbon atom (\res{Leu, Phe, Tyr,
  Trp, His}), a \res{Pro}, and two hydrophilic \res{OE1} Oxygen atoms
  (\res{Glu, Gln}) \mbox{$11.89$ \AA}\ apart.\\ 
  $[$Pos cover = 11, Neg cover = 2$]$\\
\item It contains an \res{Asp} residue $B$, two identical atoms $Q$ and
  $X$, and a hydrophilic hydrogen-bonding atom $K$.  Atoms $K$, $Q$
  and $X$ have the same charge.  $B$'s \res{OD1} Oxygen atom share the same
  Y-coordinate with $K$ and the same Z-coordinate with $Q$.  Atoms $X$ and $K$
  are \mbox{$8.29$ \AA}\ apart.\\
  $[$Pos cover = 8, Neg cover = 0$]$\\
\item It contains a \res{Ser} residue, and two \res{NE2} Nitrogen atoms
  (\res{Gln, His}) \mbox{$3.88$ \AA}\ apart.\\ 
  $[$Pos cover = 8, Neg cover = 2$]$\\
\item It contains an \res{Asn} residue and a \res{Phe, Tyr} or \res{His}
  residue, whose \res{CE1} Carbon atom is \mbox{$7.07$ \AA}\ away from a
  Calcium ion.\\
  $[$Pos cover = 5, Neg cover = 0$]$\\
\item It contains a \res{Lys} or \res{Arg}, a \res{Phe, Tyr} or \res{Arg},
  a \res{Trp}, and a Sulfate or a Phosphate ion.\\
  $[$Pos cover = 3, Neg cover = 0$]$
\end{enumerate}

Most of these rules closely reproduce current biochemical
knowledge. One in particular is novel. We will discuss rule relevance
in Section~\ref{S:alephRule}.

\section{Experimental Evaluation}
We evaluate our performance by comparing Aleph to several baseline
machine learning classifiers.

\subsection{Feature Vector Representation}
Unlike Aleph, the implemented baseline algorithms require a
constant-length feature vector input.  We change our binding-site
representation accordingly. We subdivide the binding-site sphere into
concentric shells as suggested by Bagley and Altman~\cite{Bagley}.
Nassif et al.~\cite{Nassif} subdivided the sphere into $8$ layers
centered at the binding-site centroid.  The first layer had a width of
\mbox{$3$ \AA}\ and the subsequent $7$ layers where \mbox{$1$
  \AA}\ each.  Their results show that the layers covering the first
\mbox{$5$ \AA}, the subsequent \mbox{$3$ \AA}\ and the last \mbox{$2$
  \AA}\ share several attributes.  We thereby subdivide our
binding-site sphere into $3$ concentric layers, with layer width of
\mbox{$5$ \AA}, \mbox{$3$ \AA}\ and \mbox{$2$ \AA}\ respectively.  For
each layer, our algorithm reports the total number of atoms in that
layer and the fraction of each atomic property (charge,
hydrogen-bonding, hydrophobicity).  For example, feature \lq\lq layer
$1$ hydrophobic atoms\rq\rq\ represents the fraction of the first
layer atoms that are hydrophobic.

The ILP predicate representation allows it to implicitly infer
residues from atomic names and properties.  We use a weakly expressive
form to explicitly include amino acids in the feature vector
representation.  Amino acids are categorized into subgroups, based on
their structural and chemical properties~\citep{Betts}. We base our
scheme on the representation adopted by Nassif et al.~\cite{Nassif},
grouping histidine, previously a subclass on its own, with the rest of
the aromatic residues. Histidine can have roles which are unique among
the aromatic amino acids. We group it with other aromatics because it
was not selected as a relevant feature in our previous work.  Gilis
et al.~\cite{Gilis} report the mean frequencies of the individual
amino acids in the proteomes of $35$ living organisms.  Adding up the
respective frequencies, we get the expected percentage $p$ of each
residue category.  We categorize the residue features into \lq\lq
low\rq\rq, \lq\lq normal\rq\rq\ and \lq\lq high\rq\rq.  A residue
category feature is mapped to \lq\lq normal\rq\rq\ if its percentage
is within $2 \times \sqrt{p}$ of the expected value $p$.  It is mapped
to \lq\lq low\rq\rq\ if it falls below, and to \lq\lq high\rq\rq\ if
it exceeds the cut-off.  Table~\ref{t:residue_sheme_cut_points}
accounts for the different residue categories, their expected
percentages, and their cut-off values mapping boundaries. Given a
binding site, our algorithm computes the percentage of amino acids of
each group present in the sphere, and records its nominal value.  We
ignore the concentric layers, since a single residue can span several
layers.

\begin{table}
\caption{Residue grouping scheme, expected percentage, and mapping boundaries}
\begin{tabular*}{\textwidth}{@{\extracolsep{\fill}}*{5}{l}}
  \hline\noalign{\smallskip} Residue & Amino Acids & Expected & Lower
  & Upper \\ Category & & Percentage & Bound &
  Bound\\ \noalign{\smallskip}\hline\noalign{\smallskip} Aromatic &
  \res{His, Phe, Trp, Tyr} & 10.81\% & 4.23\% & 17.39\% \\ Aliphatic &
  \res{Ala, Ile, Leu, Met, Val} & 34.19\% & 22.50\% & 45.88\%
  \\ Neutral & \res{Asn, Cys, Gln, Gly, Pro, Ser, Thr} & 31.53\% &
  20.30\% & 42.76\% \\ Acidic & \res{Asp, Glu} & 11.91\% & 5.01\% &
  18.81\% \\ Basic & \res{Arg, Lys} & 11.55\% & 4.75\% & 18.35\%
  \\ \noalign{\smallskip}\hline
\end{tabular*}
\label{t:residue_sheme_cut_points}
\end{table}

The final feature vector is a concatenation of the atomic and residue
features.  It contains the total number of atoms and the atomic
property fractions for each layer, in addition to the residue
features.  It totals $27$ continuous and $5$ nominal features.

\subsection{Baseline Classifiers}
This section details our implementation and parametrization of the
baseline algorithms.  Refer to Mitchell~\cite{Mitchell} and Duda et
al.~\cite{Duda1} for a complete description of the algorithms.

\subsubsection{$k$-Nearest Neighbor}
The scale of the data has a direct impact on $k$-Nearest Neighbor's
($k$NN) classification accuracy.  A feature with a high data mean and
small variance will a priori influence classification more than one
with a small mean and high variance, regardless of their
discrimination power~\citep{Duda1}.  In order to put equal initial
weight on the different features, the data is standardized by scaling
and centering.

Our implementation handles nominal values by mapping them to ordinal
numbers.  It uses the Euclidean distance as a distance function.  It
chooses the best $k$ via a leave-one-out tuning method.  Whenever two
or more $k$'s yield the same performance, it adopts the larger one.
If two or more examples are equally distant from the query, and all
may be the $k$th nearest neighbor, our implementation randomly
chooses.  On the other hand, if a decision output tie arises, the
query is randomly classified.

We also implement feature backward-selection (BS$k$NN) using the
steepest-ascent hill-climbing method.  For a given feature set, it
removes one feature at a time and performs $k$NN.  It adopts the trial
leading to the smaller error.  It repeats this cycle until removing
any additional feature increases the error rate.  This
implementation is biased towards shorter feature sets, going by
Occam's razor principle.

\subsubsection{Naive Bayes}
Our Naive Bayes (NB) implementation uses a Laplacian smoothing
function.  It assumes that continuous features, for each output class,
follow the Gaussian distribution.  Let $X$ be a continuous feature to
classify and $Y$ the class.  To compute $P(X|Y)$, it first calculates
the normal $z$-score of $X$ given $Y$ using the $Y$-training set's
mean $\mu_Y$ and standard deviation $s_Y$: $z_Y = (x - \mu_Y)/s_Y$.
It then converts the $z$-score into a $[0,1]$ number by integrating
the portions of the normal curve that lie outside $\pm z$. We use this
number to approximate $P(X|Y)$. This method returns $1$ if $X = \mu$,
and decreases as $X$ steps away from $\mu$:
\begin{equation}
  P(X|Y) = \int_{|z_Y|}^\infty normalCurve + \int_{-|z_Y|}^{-\infty} normalCurve\,.
\end{equation}

\subsubsection{Decision Trees}
Our Decision Tree implementation uses information gain as a
measure for the effectiveness of a feature in classifying the training
data.  We incorporate continuous features by dynamically defining new
discrete-valued attributes that partition the continuous attribute
value into a discrete set of intervals.  We prune the resulting tree
using a tuning set.  We report the results of both pruned (Pr DT) and
unpruned decision trees (DT).

\subsubsection{Perceptron}
Our perceptron (Per) implementation uses linear units and performs a
stochastic gradient descent.  It is therefore similar to a logistic
regression. It automatically adjusts the learning rate, treats the
threshold as another weight, and uses a tuning set for early stopping
to prevent overfitting.  We limit our runs to a maximum of $1000$
epochs.

\subsubsection{Sequential Covering}
Sequential Covering (SC) is a propositional rules learner that returns
a set of disjunctive rules covering a subset of the positive examples.
Our implementation uses a greedy approach.  It starts from the empty
set and greedily adds the best attribute that improves rule
performance.  It discretizes continuous attributes using the same
method as Decision Trees.  It sets the rule coverage threshold to $4$
positive examples and no negative examples.  The best attribute to add
is the one maximizing:
\begin{equation}
|entropy(parent) - entropy(child)| * numberOfPositives(child)\,.
\end{equation}

\subsection{Baseline Classifiers Results}\label{S:baseline}

We apply the same $10$-folds cross-validation to Aleph and all the
baseline classifiers.  Table~\ref{t:baseline} tabulates the error
percentage per testing fold, the mean, standard deviation and the
$95\%$ level confidence interval for each classifier.

\begin{table}
\caption{$10$-folds cross-validation test error percentage, mean,
  standard deviation and the $95\%$ level confidence interval for the
  baseline algorithms and Aleph}
\label{t:baseline}
\begin{tabular*}{\textwidth}{@{\extracolsep{\fill}}*{9}{l}}
  \hline\noalign{\smallskip}
  Fold & $k$NN & BS$k$NN & NB & DT &Pr DT& Per & SC & Aleph\\
  \noalign{\smallskip}\hline\noalign{\smallskip}
  0 & $25.0$  & $25.0$ & $43.75$ & $31.25$ & $37.5$ & $43.75$ &$31.25$ & $25.0$\\
  1 & $25.0$  & $25.0$  & $25.0$ & $31.25$ & $25.0$ & $43.75$ &$31.25$ & $37.5$\\
  2 & $18.75$ & $18.75$ & $25.0$ & $12.5$ & $25.0$ & $25.0$ & $25.0$ & $25.0$\\
  3 & $18.75$ & $18.75$ & $37.5$ &$6.25$ & $12.5$ & $31.25$ & $12.5$& $50.0$\\
  4 & $25.0$ &  $37.5$  & $37.5$ & $25.0$ & $37.5$ & $25.0$ & $12.5$ & $31.25$\\
  5 & $31.25$ & $31.25$ & $37.5$ & $31.25$ &$18.75$ &$37.5$ & $31.25$& $18.75$\\
  6 & $31.25$ & $18.75$ & $25.0$ & $37.5$ & $31.25$ & $37.5$ & $25.0$& $25.0$\\
  7 & $31.25$ & $25.0$  & $37.5$ & $25.0$ & $31.25$ & $31.25$ &$37.5$& $43.75$\\
  8 & $18.75$ & $18.75$ & $31.25$ & $25.0$ & $12.5$ & $31.25$ & $31.25$& $25.0$\\
  9 & $31.25$ & $31.25$ & $50.0$ &$50.0$ & $31.25$ & $43.75$ &$25.0$& $43.75$\\
  \noalign{\smallskip}\hline\noalign{\smallskip}
  mean & $25.63$ & $25.0$ & $35.0$ & $27.5$ & $26.25$ & $35.0$ & $26.25$ & $32.5$\\
  standard deviation & $5.47$ & $6.59$ & $8.44$ & $12.22$ & $9.22$ & $7.34$ & $8.23$ & $10.54$\\
  lower bound & $21.71$ & $20.29$ & $28.97$ & $18.77$ & $19.66$ & $29.76$ & $20.37$ & $24.97$\\
  upper bound & $29.54$ & $29.71$ & $41.03$ & $36.23$ & $32.84$ & $40.24$ & $32.13$ & $40.03$\\
 \noalign{\smallskip}\hline
\end{tabular*}
\end{table}

Our SC implementation learns a propositional rule that covers at least
$4$ positive and no negative examples. It then removes all positive
examples covered by the learned rule. It repeats the process using the
remaining positive examples.  Running SC over the whole data set
generates the following rules, sorted by coverage.  Together they
cover $63$ positives out of $80$.  A site is hexose-binding if any of
these rules apply:

\begin{enumerate}
\item
  \begin{description}
  \item[If] layer $1$ negatively charged atoms density $> 0.0755$
  \item [and] layer $2$ positively charged atoms density $< 0.0155$
  \item [and] layer $3$ negatively charged atoms density $> 0.0125$
  \item $[$Pos cover = 32$]$\\
  \end{description}
\item
  \begin{description}
  \item[If] layer $1$ non hydrogen-bonding atoms density $< 0.559$
  \item [and] layer $1$ hydrophobic atoms density $> 0.218$
  \item [and] layer $3$ hydrophilic atoms density $> 0.3945$
  \item $[$Pos cover = 14$]$\\
  \end{description}
\item
  \begin{description}
  \item[If] layer $1$ negatively charged atoms density $> 0.0665$
  \item [and] layer $1$ hydroneutral atoms density $< 0.2615$
  \item [and] layer $1$ non hydrogen-bonding atoms density $> 0.3375$
  \item [and] layer $3$ atoms number $< 108.5$
  \item $[$Pos cover = 12$]$\\
  \end{description}
\item
  \begin{description}
  \item[If] layer $1$ negatively charged atoms density $> 0.0665$
  \item [and] layer $2$ atoms number $> 85.5$
  \item [and] layer $1$ negatively charged atoms density $< 0.3485$
  \item $[$Pos cover = 5$]$
  \end{description}
\end{enumerate}

\section{Discussion} 
Despite its average performance, the main advantage of ILP is the
insight it provides to the underlying discrimination process.

\subsection{Aleph's Performance}
Aleph's error rate of $32.5\%$ has a $p$-value $< 0.0002$, according
to a two-sided binomial test.  Random guessing would return $50\%$,
since the number of positives and negatives are equal.  According to a
paired t-test at the $95\%$ confidence level, the difference between
Aleph and each of the baseline algorithms is not statistically
significant.  Aleph's mean error rate ($32.5\%$) and standard
deviation ($10.54\%$) are within the ranges observed for the baseline
classifiers, $[25\%,35\%]$ and $[5.47\%,12.22\%]$ respectively (see
Table~\ref{t:baseline}).

Aleph's error rate is also comparable to other general sugar binding
site classifiers, ranging from $31\%$ to $39\%$, although each was run
on a different data set (Table~\ref{t:compare}).  On the other hand,
specific sugar binding sites classifiers have a much better
performance (Table~\ref{t:compare}).  COTRAN~\citep{SUJ1}
galactose-binding site classifier achieves a $5.09\%$ error while
Nassif et al.~\cite{Nassif} glucose-binding site classifier reports an
error of $8.11\%$.  This may suggest that the problem of recognizing
specific sugars is easier than classifying a family of sugars.

\begin{table}
  \caption{Error rates achieved by general and specific sugar binding site
    classifiers.  Not meant as a direct comparison since the data sets
    are different.}
  \label{t:compare}
    \begin{tabular*}{\textwidth}{@{\extracolsep{\fill}}llp{.45\linewidth}}
      \hline\noalign{\smallskip}
      Program   & Error ($\%$) & Method and Data set\\
      \noalign{\smallskip}\hline\noalign{\smallskip}
      \multicolumn{3}{l}{General sugar binding sites classifiers}\\
      \noalign{\smallskip}
      ILP hexose predictor & $32.50$ & $10$-folds cross-validation, $80$ hexose
      and $80$ non-hexose or non-binding sites\\
      Shionyu-Mitsuyama et al.~\cite{Shionyu} & $31.00$ & Test set, $61$ polysaccharide binding sites\\
      Taroni et al.~\cite{Taroni} & $35.00$ & Test set, $40$ carbohydrate binding sites\\
      Malik and Ahmad~\cite{Malik} & $39.00$ & Leave-one-out, $40$ carbohydrate and $116$ non-carbohydrate binding sites\\
      \noalign{\medskip}
      \multicolumn{3}{l}{Specific sugar binding sites classifiers}\\
      \noalign{\smallskip}
      COTRAN~\citep{SUJ1} & $5.09$ & Overall performance over $6$-folds, totaling $106$ galactose and $660$ non-galactose binding sites\\
      Nassif et al.~\cite{Nassif} & $8.11$ & Leave-one-out, $29$ glucose and $35$ non-glucose or non-binding sites\\
      \noalign{\smallskip}\hline
    \end{tabular*}
\end{table}

\subsection{Aleph Rules Interpretation}\label{S:alephRule}
Contrary to black-box classifiers, ILP provides a number of
interesting insights.  It infers most of the established biochemical
information about residues and relations just from the PDB-atom names
and properties. We hereby interpret Aleph's rules detailed in
Section~\ref{S:AlephResuts}.

Rules $1$, $2$, $5$, $8$ and $9$, rely on the aromatic residues
\res{Trp, Tyr} and \res{Phe}.  This highlights the docking interaction
between the hexose and the aromatic residues~\citep{Malik,SUJ1,SUJ2}.
The aromatic residues stack against the apolar sugar pyranose ring
which stabilizes the bound hexose.  \res{His} is mentioned in many of
the rules, along-side other aromatics ($5$, $8$) or on its own ($4$,
$7$). Histidine provides a similar docking mechanism to \res{Trp, Tyr}
and \res{Phe}~\citep{Quiocho}.

All nine rules require the presence of a planar polar residue
(\res{Asn, Asp, Gln, Glu, Arg}).  These residues have been identified
as the most frequently involved in the hydrogen-bonding of
hexoses~\citep{Quiocho}.  The hydrogen bond is probably the most
relevant interaction in protein binding in general.

Rules $1$, $2$, $3$, $5$ and $6$ call for acidic residues with a
negative partial charge (\res{Asp, Glu}), or for atoms with a negative
partial charge.  The relative high negative density observed may be
explained by the dense hydrogen-bond network formed by the hexose
hydroxyl groups.

Some rules require hydrophobic atoms and residues, while others require
hydrophilic ones.  Rule $5$ requires both and reflects the dual nature
of sugar docking, composed of a polar-hydrophilic aspect establishing
hydrogen bonds and a hydrophobic aspect responsible for the pyranose
ring stacking~\citep{Taroni}.

A high residue-sugar propensity value reflects a high tendency of that
residue to be involved in sugar binding.  The residues having high
propensity values are the aromatic residues, including histidine, and
the planar polar residues~\citep{Taroni}.  This fact is reflected by
the recurrence of high propensity residues in all rules.

Rules $8$ and $9$ require, and rule $1$ is satisfied by, the presence
of different ions (Calcium, Sulfate, Phosphate), confirming the
relevance of ions in hexose binding~\citep{Quiocho}.

Rule $6$ specifies a triangular conformation of three atoms
within the binding-site. This highlights the relevance of the
binding-site's spatial features.  On the other hand, we note the
absence of the site's centroid literal from the resulting Aleph rules.
The center is merely a geometric parameter and does not have any
functional role.  In fact, the binding site center feature was not
used in most computational classifying approaches.  Taroni et
al.~\cite{Taroni} and Malik and Ahmad~\cite{Malik} ignore it,
Shionyu-Mitsuyama et al.~\cite{Shionyu} use the pyranose ring $C3$
atom instead, and Nassif et al.~\cite{Nassif} indirectly use it to
normalize distances from atoms within the binding pocket to atoms in
the ligand.  Only Sujatha and Balaji~\cite{SUJ1} explicitly refer to
the center.  These results confirm that the biochemical composition of
the binding-site and the relative $3$-dimensional positioning of its
atoms play a much more important role in substrate specificity than
the exact location of the ligand's center.

Rao et al.~\citep{Rao1} report a dependency between \res{Phe/Tyr} and
\res{Asn/Asp} in the Lectin protein family. This dependency is
reflected in rules $2$ and $8$.  Similarly, rule $1$ suggests a
dependency between \res{Trp} and \res{Glu}, a link not previously
identified in literature. This novel relationship merits further
investigation and highlights the rule-discovery potential of ILP.

\subsection{Baseline Algorithms Insight}
In addition to providing a basis for comparison with Aleph, the
baseline algorithms shed additional light on our data set and hexose
binding site properties.

Naive Bayes and Perceptron return the highest mean error rates,
$35.0\%$.  Naive Bayes is based on the simplifying assumption that the
attribute values are conditionally independent given the target value.
This assumption does not hold for our data. In fact, charge,
hydrogen-bonding and hydrophobicity values are correlated. Like Naive
Bayes, Perceptron correctly classifies linearly separable data.  Its
high error stresses the fact that our data is not linearly separable.

On the other hand, backward-selection $k$NN algorithm, with the lowest
mean error rate ($25.0\%$), outperforms all other classifiers. This
provides further evidence that similar sites, in terms of biochemical
and spatial properties, are good predictors of
binding~\cite{Gold}. $k$NN's good performance further highlights both
the correlation between our features, and the data's non-linearity.

\medskip
Like Aleph, Sequential Covering's rules provide insight into the
discriminating process.  Unlike Aleph's first-order logic rules,
propositional rules are less expressive and reflect a smaller number
of biochemical properties.  We hereby interpret SC's rules detailed in
Section~\ref{S:baseline}.

Although SC uses an explicit representation of residues, it completely
ignores them in the retained rules. Only atomic biochemical features
influence the prediction.  This may be due to the fact that it is the
binding-site's atoms, rather than overall residues, that bind to and
stabilize the docked hexose.  These atoms may not be mapped to
conserved specific residues.

Another general finding is that most rule antecedents are layer $1$
features.  This reflects the importance of the atoms lining the
binding-site, which establish direct contact with the docking hexose.
Layers $2$ and $3$ are farther away and hence have weaker interaction
forces.

Only four amino acid atoms have a partial negative charge in our
representation, in addition to the infrequent Sulfate and Phosphate
Oxygens~\cite{Nassif}. The first rule, covering most of the positive
examples, clearly suggests a binding-site with a high density of
negatively charged atoms. The first and third antecedents explicitly
specify layers with a negatively charged atomic density above some
thresholds.  The second one implicitly states so by opting for a
non-positively charged layer.  The relative high negative density
observed may be explained by the dense hydrogen-bond network formed by
the hexose hydroxyl groups~\citep{Quiocho}.  The fourth rule is
similar.  It imposes a bond on the first layer's negative charge,
between $0.0665$ and $0.3485$.  Although it is well established in the
literature that hexose forms hydrogen bonds through its hydrogen's
partial positive charge~\citep{Zhang}, the binding site itself is not
known to be negatively charged.  Rule $4$ captures this distinction,
requiring the binding site to have an above-average density of
negatively charged atoms, while still setting upper-bounds.

The second rule requires a high density of hydrophobic atoms in layer
$1$, and a high density of hydrophilic atoms in layer $3$.  This
reflects the dual hydrophobic-hydrophilic nature of hexose
binding~\citep{Taroni}.  It also indirectly specifies a high
hydrogen-bonding density by imposing an upper limit for the
non-hydrogen bonding atoms.

The third rule is a combination of the other ones.  First it demands a
slightly negative first layer.  Second, it requires hydroneutral
atoms.  Third it implicitly asks for hydrogen bonding atoms by setting
a relatively low threshold for the much more abundant non-hydrogen
bonding atoms. It is worth to note that the number of atoms in the
third layer is capped by $108$. This may reflect a particular
spatial-arrangement of atoms in hexose binding sites.

\section{Conclusion}
In this work, we present the first attempt to model and predict hexose
binding sites using ILP.  We investigate the empirical support for
biochemical findings by comparing Aleph induced rules to actual
biochemical results.  Our ILP system achieves a similar accuracy as
other general protein-sugar binding sites black-box classifiers, while
offering insight into the discriminating process.  With no prior
biochemical knowledge, Aleph was able to induce most of the known
hexose-protein interaction biochemical rules, with a performance that
is not significantly different than several baseline algorithms.  In
addition, ILP finds a previously unreported dependency between
\res{Trp} and \res{Glu}, a novel relationship that merits further
investigation.

\subsubsection*{Acknowledgments.}
This work was partially supported by US National Institute of Health
(NIH) grant R01CA127379-01.  We would like to thank Jose Santos
for his inquiries that led us to improve our Aleph representation.

\bibliographystyle{splncs03}
\bibliography{XBib}

\end{document}